\documentclass[aps,prl,onecolumn,superscriptaddress]{revtex4-1}

\usepackage{graphicx}
\usepackage[flushleft]{caption2}

\begin{document}
\title{Edge Temperature Ring Oscillation Modulated by Turbulence Transition for Sustaining Stationary Improved Energy Confinement Plasmas}

\author{A. D. Liu}
\email{lad@ustc.edu.cn}
\affiliation{University of Science and Technology of China, Hefei, Anhui 230026, China}
\author{X. L. Zou}
\affiliation{CEA, IRFM, F-13108 St Paul Les Durance, France} 
\author{M. K. Han}
\affiliation{Key Laboratory of Materials Modification by Beams, School of Physics, Dalian University of Technology, Dalian 116024, China}
\affiliation{Southwestern Institute of Physics, Chengdu 610041, China}
\author{T. B. Wang}
\affiliation{Southwestern Institute of Physics, Chengdu 610041, China}
\affiliation{Department of Applied Physics, Ghent University, B-9000 Ghent, Belgium}
\author{C. Zhou}
\affiliation{University of Science and Technology of China, Hefei, Anhui 230026, China}
\author{M. Y. Wang}
\email{wmyuan@mail.ustc.edu.cn}
\affiliation{University of Science and Technology of China, Hefei, Anhui 230026, China}
\affiliation{Hebei Key Laboratory of Compact Fusion, Langfang 065001, China}
\author{Y. M. Duan}
\affiliation{Institute of Plasma Physics, Chinese Academy of Sciences, Hefei 230031, China}
\author{G. Verdoolaege}
\affiliation{Department of Applied Physics, Ghent University, B-9000 Ghent, Belgium}
\author{J. Q. Dong}
\affiliation{Southwestern Institute of Physics, Chengdu 610041, China}
\author{Z. X. Wang}
\affiliation{Key Laboratory of Materials Modification by Beams, School of Physics, Dalian University of Technology, Dalian 116024, China}
\author{X. Feng}
\author{J. L. Xie}
\author{G. Zhuang}
\author{W. X. Ding}
\affiliation{University of Science and Technology of China, Hefei, Anhui 230026, China}
\author{S. B. Zhang}
\author{Y. Liu}
\author{H. Q. Liu}
\author{L. Wang}
\author{Y. Y. Li}
\author{Y. M. Wang}
\author{B. Lv}
\author{G. H. Hu}
\author{Q. Zhang}
\author{S. X. Wang}
\author{H. L. Zhao}
\affiliation{Institute of Plasma Physics, Chinese Academy of Sciences, Hefei 230031, China}
\author{C. M. Qu}
\author{Z. X. Liu}
\author{Z. Y. Liu}
\author{J. Zhang}
\author{J. X. Ji}
\author{X. M. Zhong}
\author{T. Lan}
\author{H. Li}
\author{W. Z. Mao}
\author{W. D. Liu}
 \affiliation{University of Science and Technology of China, Hefei, Anhui 230026, China}
\author{EAST Team}

\date{\today}

\begin{abstract}
A reproducible stationary improved confinement mode (I-mode) has been achieved recently in the Experimental Advanced Superconducting Tokamak, featuring good confinement without particle transport barrier, which could be beneficial to solving the heat flux problem caused by edge localized modes (ELM) and the helium ash problem for future fusion reactors. The microscopic mechanism of sustaining stationary I-mode, based on the coupling between turbulence transition and the edge temperature oscillation, has been discovered for the first time. A radially localized edge temperature ring oscillation (ETRO) with azimuthally symmetric structure ($n=0$,$m=0$) has been identified and it is caused by alternative turbulence transitions between ion temperature gradient modes (ITG) and trapped electron modes (TEM). The ITG-TEM transition is controlled by local electron temperature gradient and consistent with the gyrokinetic simulations. The self-organizing system consisting with ETRO, turbulence and transport transitions plays the key role in sustaining the I-mode confinement. These results provide a novel physics basis for accessing, maintaining and controlling stationary I-mode in the future.
\end{abstract}


\maketitle
Simultaneous establishment of both high confinement and steady state is one of the most crucial goals for magnetic fusion research \cite{Loarte2007NF,Kikuchi2012RMP}. Although the high-confinement mode (H-mode) \cite{Wagner1982PRL}, which is characterized by the edge transport barriers of both the density and temperature profiles, is already considered as the baseline operation scenario for the International Tokamak Experimental Reactor project (ITER), unacceptable heat flux to the divertor targets caused by large  edge localized modes (ELMs) due to relaxation of the edge transport barrier is still one of the most crucial issues in fusion research \cite{Loarte2003JNM}. An alternative improved confinement regime (I-mode) \cite{Whyte2010NF}, which features good confinements comparable to H-mode, absence of ELM, and no significant reduction in particle transport, could be a possible solution and has been widely studied on various divertor Tokamaks \cite{Hubbard2016NF,Ryter2017NF}. Moreover, I-mode could also be an ascendant candidate to transport helium ash out of a burning plasma for future fusion reactor due to the absence of particle transport barrier \cite{Reiter1991PPCF}. However, although it is widely accepted that the I-mode edge should be ideal magnetohydrodynamic (MHD) stable and dominated by turbulence transport, the intrinsic microscopic mechanism to sustain such specific improved confinement is still unclear \cite{Hubbard2016NF,Manz2015NF,Marinoni2015NF}.


Recently stationary I-mode has been identified in the Experimental Advanced Superconducting Tokamak (EAST) \cite{Feng2019NF}, with the common characteristics reported on other devices: unfavorable $\vec{B}\times\nabla B$ drift direction, weakly coherence mode (WCM) with the frequency range of $40-150kHz$ and no particle transport barrier at the edge plasmas. The most noteworthy feature is that I-mode on EAST could sustain a long time without transition to H-mode, the maintenance time could reach several seconds and only limited by the auxiliary heating duration. It was found that the stationary I-mode has no auxiliary heating preference and is always accompanied by a low-frequency coherent mode of $6-12 kHz$. This mode could be observed by most edge plasma diagnostics, such as magnetic coils, divertor probes, $D_{\alpha}$ filterscopes, electron cyclotron emissions (ECE), bolometers and so on. Similar low-frequency oscillations have also been reported on ASDEX Upgrade and Alcator C-Mod \cite{Cziegler2013POP,Manz2015NF} and the common mode should play important role in turbulence transport process at the edge region. In tokamak plasmas, both theories and experiments indicate that the turbulent transport is mainly attributed to two electrostatic turbulence: ion temperature gradient modes (ITG) \cite{Romanelli1989PFB} and trapped electron modes (TEM) \cite{Weiland1992NF,Horton1999RMP}, and they propagate in opposite directions in the plasma frame and lead to opposite convective velocities \cite{Garbet2003PRL,Zhong2013PRL}. Experiments also suggested that ITG-TEM transition should be responsible for the spontaneous internal transport barrier formation \cite{Xiao2010PRL}, as well as the Ohmic L-mode energy confinement changing from the linear to the saturated regime \cite{Rice2011PRL}.

In this letter, it has been demonstrated that the common low frequency coherent mode is actually a radially localized edge temperature ring oscillation (ETRO) with azimuthally symmetric structure, and it is caused by the alternative transitions between TEM and ITG. The experimental findings revealed a novel perspective on the stationary I-mode microscopic mechanism.

I-mode experiments were carried out in the EAST tokamak with plasma major radius $R=1.9 m$ and plasma minor radius $a=0.45 m$ \cite{Li2013NP}. The electron temperature $T_e$ profile is measured by $32$-channel ECE and Thomson scattering (TS) system. $2D$ evolution of $T_e$ in the cross section could be measured by a $384$ channel ($24$ vertical $\times16$ radial) electron cyclotron emission imaging system (ECEI) \cite{Gao2018RSI}. Based on the relative self-dependent calibration technique \cite{Han2018RSI}, the absolute fluctuation amplitude at specified frequency range of $T_e$ could also be estimated. Plasma radiation is measured by bolometer arrays with $64$ chords in the whole cross section \cite{Duan2011PST}, which will be used for the radiation reconstruction later. The turbulences at eight different radial locations are simultaneously measured through an eight-channel Doppler reflectometry (DR) \cite{Hu2017RSI}. The measured turbulence wavenumber is $k_{\perp}=4-6cm^{-1}$. The Doppler shift $\omega_{d}$ of the turbulence could be written as $\omega_{d}=k_{\perp}\upsilon_{\perp}$, among which $\upsilon_{\perp}$ is the turbulence rotation velocity in the perpendicular direction to the magnetic field, and $\vec{\upsilon}_{\perp}=\vec{\upsilon}_{E\times B}+\vec{\upsilon}_{pha}$, where $\vec{\upsilon}_{E\times B}$ is the plasma $E\times B$ rotation due to radial electrical field and $\vec{\upsilon}_{pha}$ is the turbulence phase velocity. For simplicity, the name ETRO will be used at the beginning before detailed analyses.

\begin{figure}
\includegraphics[width=8.5cm,clip]{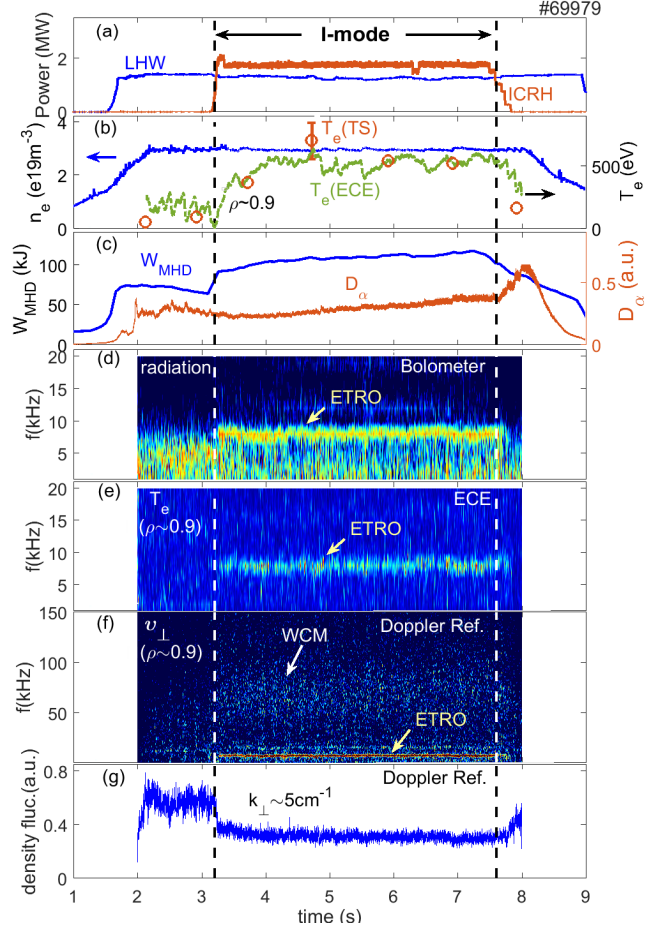}
\caption{\label{f1} From top to bottom, temporal evolutions of (a) auxiliary heating power, (b) chord averaged density and edge $T_e$, (c) plasma stored energy $W_{MHD}$ and $D_{\alpha}$ signal; time-frequency spectra of (d) the bolometer signal, (e) the ECE signal at $\rho\sim0.9$ (f) and turbulence rotation velocity $\upsilon_{\perp}$; (g) turbulence intensity evolution ($k_{\perp}=5cm^{-1}$) with I-mode from $t=3.2s$ to $t=7.5s$.}
\end{figure}

Figure \ref{f1} shows a typical stationary I-mode discharge lasting about $4.3s$ with plasma current $Ip=550kA$ and toroidal magnetic field $B_t=2.2T$. The auxiliary heating included $1.2MW$ low hybrid wave (LHW) and $1.8MW$ ion cyclotron resonance heating (ICRH). ICRH triggered I-mode from $t=3.2s$ to $t=7.5s$, as shown in panel(a). The chord averaged density and $T_e$ at $\rho\sim0.9$ are displayed in panel(b), suggesting that during L-I transition the density is almost unchanged while $T_e$ increased abruptly. The plasma stored energy $W_{MHD}$ is increased to a new plateau while no significant change is observed in the $D_{\alpha}$ signal during the transition, as shown in panel(c). The time-frequency spectra of the bolometer signal and the ECE signal at $\rho\sim0.9$ are displayed in panels (d) and (e) respectively, where the ETRO with $f=8-9kHz$ is obvious during the whole I-mode, suggesting that the oscillation may be dominate by $T_e$ perturbation. The time-frequency spectra of turbulence rotation velocity $\upsilon_{\perp}$ is shown in panel(f), where the ETRO and WCM could both be obviously distinguished during I-mode. Panel(g) displayed the time evolution of turbulence intensity with $k_{\perp}\sim5cm^{-1}$, which is abruptly dropped by half during the L-I transition, due to the temperature pedestal formation and the increased $E_r$ shearing at the edge region \cite{Viezzer2013NF,Feng2019NF}.

\begin{figure}
	\includegraphics[width=8.5cm,clip]{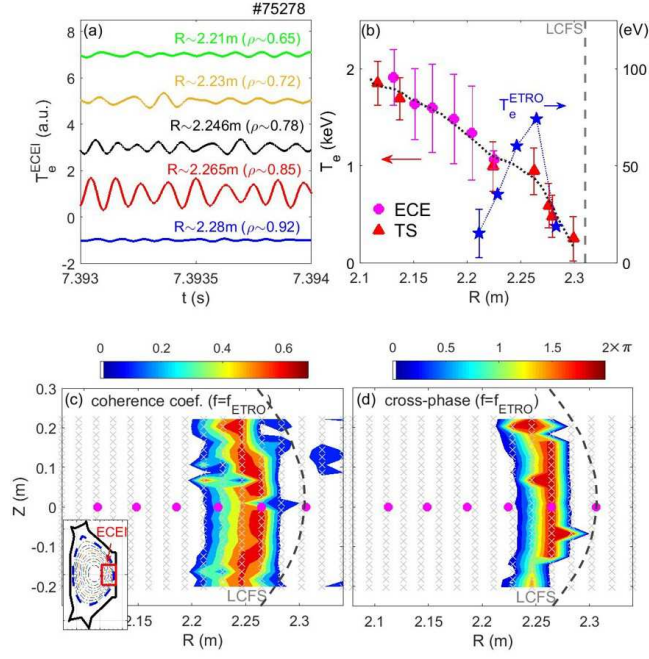}
	\caption{\label{f3} (a) The calibrated data from $5$ ECEI channels at ETRO frequency. (b) $T_e$ profile combined from ECE and TS diagnostics during I-mode, as well as the distribution of $\tilde{T}^{ETRO}_{e}$ rms values. Contour plots of (c) correlation coefficients and (d) cross-phases at ETRO frequency between all ECEI channels and the fourth ECE channel. The ECEI measuring range is shown in the lower left corner.}
\end{figure}

ECEI was moved to the plasma edge region at I-mode shot $\#75278$ to measure the $T_e$ poloidal structure. The ECEI data from $5$ channels at the equatorial plane filtered at ETRO frequency are displayed in fig.\ref{f3}(a), showing that the strongest electron temperature perturbation at ETRO frequency $\tilde{T}^{ETRO}_e$ appeared at $R\sim2.265m$($\rho\sim0.85$), and then radially decayed both outward and inward with much stronger outward decay. The radial distribution of absolute perturbation amplitude of ETRO, as well as the $T_e$ profile combined with ECE and TS diagnostics, are both displayed in fig.\ref{f3}(b). The root-mean-square (rms) of strongest $\tilde{T}^{ETRO}_e$ is about $78eV$ with relative fluctuation amplitude of $\tilde{T}^{ETRO}_e/T_e\sim9\%$. 

The fourth ECE channel has the strongest ETRO amplitude among all ECE channels and was used as a reference, the coherences between it and all ECEI signals could show that how $\tilde{T}^{ETRO}_e$ distributed on the cross-section. Fig.\ref{f3}(c) and (d) display the contour plots of correlation coefficients and cross-phases at ETRO frequency respectively. It could be found that the maximum coherent coefficient is $\sim0.7$ at $(R,Z)\sim(2.26m,0)$ with cross-phase close to $2\pi$. Several conclusions could be deduced: firstly, the coefficients and cross-phases presented distinct uniform structures along the flux surfaces, implying the poloidal symmetry of $\tilde{T}^{ETRO}_e$; secondly, the radial distribution of coefficients suggested that the outward decay length is much smaller than the inward decay length, consistent with the amplitude distribution in fig.\ref{f3}(b); thirdly, it should be noted that ECE system \cite{Liu2018FED} is not toroidally located at the ECEI port \cite{Gao2018RSI}, and the cross-phase of $2\pi$ is from two $\tilde{T}^{ETRO}_e$ separated with $3/4\pi$ toroidal angle, so the in-phase result is consistent with the toroidal symmetry estimated through $16$ groups of toroidal magnetic probes.

\begin{figure}
	\includegraphics[width=8.5cm,clip]{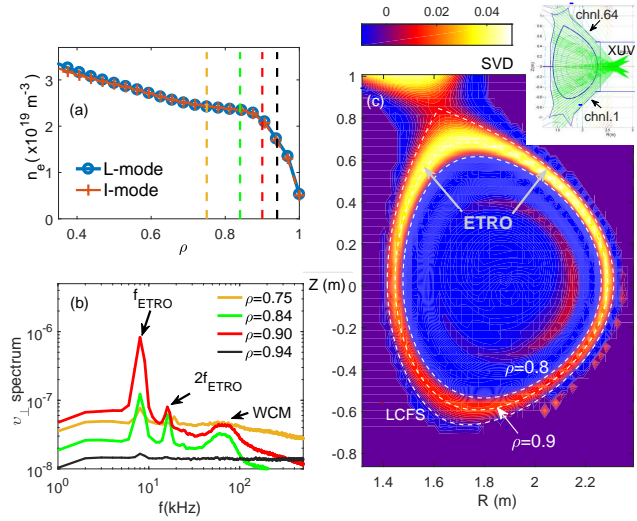}
	\caption{\label{f2} (a) The density profiles at $t=3s$ and $t=6s$ in shot $\#69979$, and four DR measurement positions. (b) the spectra of turbulence rotation velocity $\upsilon_{\perp}$ during I-mode. (c) the tomography reconstruction of second SVD. Measurement chords of the 64-channel bolometer array are zoomed out in the upper right corner.}
\end{figure}

Fig.\ref{f1}(d) shows that ETRO is distinct on the bolometer signal, so another method to confirm the poloidal symmetry of ETRO is tomography reconstruction through the $64$-channel bolometer arrays across the poloidal cross-section \cite{Duan2011PST}. Fig.\ref{f2}(c) gives the second topos component of singular value decomposition (SVD) result after reconstruction delivered by Gaussian process tomography algorithm \cite{Wang2018RSI}, which represents the spatial structure of the most important perturbation during the observation. The SVD results clearly indicates an in-phase ring-like structure localized around $\rho=0.9$ with a small up-down amplitude asymmetry. The characteristic frequency of the second SVD topos component consists with the ETRO frequency and the radial location of the ring structure is also agreed with ECEI and ECE results. Fig.\ref{f2}(a) displayed the density profiles during L-mode and I-mode, and the measurement positions of four DR channels, showing that DR measurement positions are almost unchanged during L-I transition. Fig.\ref{f2}(b) displayed the four $\upsilon_{\perp}$ spectra during I-mode, showing that ETRO is strongest at $\rho=0.9$, and then quickly decayed outward, again consistent with the ECEI results in Fig.\ref{f3}. The second harmonics of ETRO frequency is also distinct on $\upsilon_{\perp}$ spectra, which will be explained later.

Through the cross-checkings between various diagnostics, it could be deduced that the low frequency coherent mode is mainly an azimuthally symmetric $T_{e}$ perturbation localized around the temperature pedestal top, with the radial range of about $4-5cm$, so it was named as edge temperature ring oscillation (ETRO). ETRO decayed outward quickly and would certainly produce the most significant electron temperature gradient ($\nabla T_e$) change at the outside location of $T_e$ pedestal top. How the local turbulence is changed with such $\nabla T_e$ oscillation? The turbulence spectra from DR will be analyzed in the following.

\begin{figure}
	\includegraphics[width=8.5cm,clip]{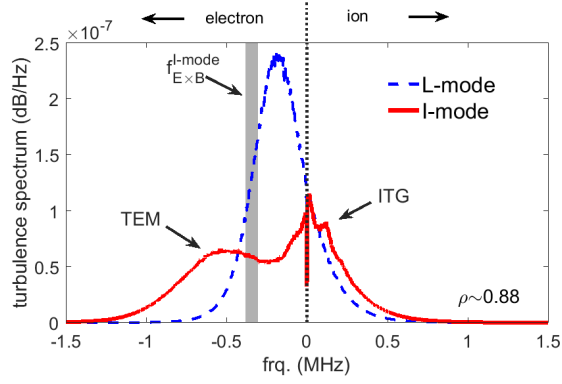}
	\caption{\label{f4} The turbulence spectra during L-mode and I-mode from DR channel at strongest ETRO location.}
\end{figure}

The turbulence intensity and rotation direction could be obtained from DR spectrum according to the Doppler shift $\omega_{d}$ and spectral area of the asymmetric peak \cite{Zou1999EPS,Holzhauer1998PPCF}. The change of turbulence spectra during L-I transition at the strongest ETRO location are shown in fig.\ref{f4}. It could be found that during L-mode, the spectrum has one Doppler shift peak at $-200kHz$, which is mainly caused by the $E\times B$ rotation \cite{Zhou2013RSI}. While during I-mode, two peaks could be found, one at $\sim-550kHz$ in the electron diamagnetic direction and the other at $\sim100kHz$ in the ion diamagnetic direction. Assuming that the ion temperature gradient is comparable to the electron temperature gradient \cite{Viezzer2013NF}, the Doppler shift due to $E\times B$ during I-mode could be roughly estimated as $-308\sim-380kHz$, as marked in fig.\ref{f4}, suggesting that the two opposite Doppler shifts should be caused by two turbulence modes with opposite phase velocities. Considering that the detected turbulence wavenumber is $k_{\perp}=4-5cm^{-1}$ and $L_{n_e}/L_{T_e}$ increased from $\sim1$ to $\sim3$ during L-I transition, the peak around $-550kHz$  is probably caused by TEM and the peak around $100kHz$ is probably caused by ITG \cite{Nilsson1995NF,Ryter2001PPCF}.

\begin{figure}
	\includegraphics[width=7.5cm,clip]{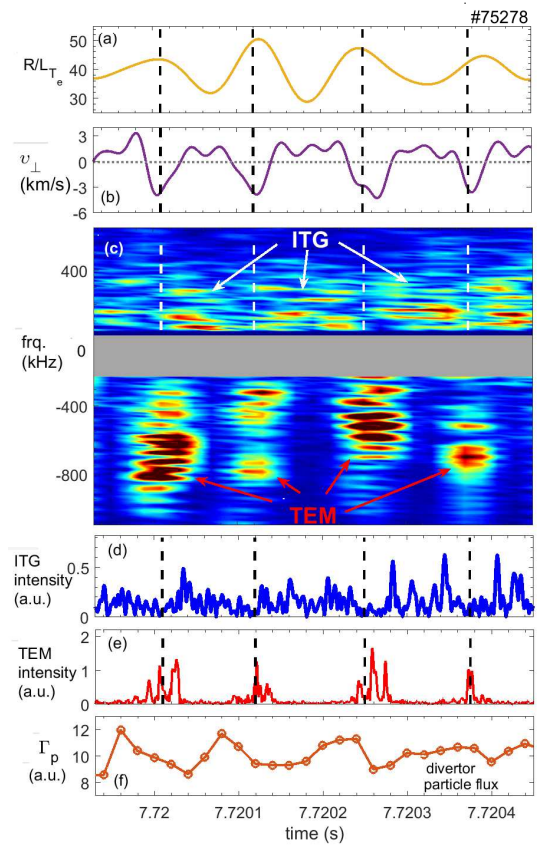}
	\caption{\label{f5} From top to bottom, temporal evolutions of (a) $R/L_{T_e}$, (b) $\upsilon_{\perp}$, (c) turbulence spectrum, (d) ITG intensity, (e) TEM intensity at strongest ETRO location, and (f) particle flux $\Gamma_p$ measured by divertor probe at the strike point during $0.5ms$ I-mode in shot $\#75278$.}
\end{figure}

Fig.\ref{f5} shows the evolutions of normalized gradient scale $R/L_{T_e}$, $\upsilon_{\perp}$, turbulence intensity and particle flux $\Gamma_p$ from the divertor probes during several ETRO periods. Panel(a) shows $R/L_{T_e}$ at the strongest ETRO location calculated from the fitting temperature profile and the ETRO evolutaions displayed in Figs.\ref{f2}(a) and (b). The averaged $R/L_{T_e}$ is about $37.8$ with $10\%-30\%$ relative change due to ETRO. Panel(b) shows that when $R/L_{T_e}$ is near the positive extreme, $\upsilon_{\perp}$ is close to the negative extreme and vice versa. Panel(c) shows the evolution of turbulence spectrum without the frequency components around zero dominated by noises and forward scatterings. It could be found that ITG and TEM are generated alternately, not simultaneously. TEM appeared when $R/L_{T_e}$ increasing while ITG appeared when $R/L_{T_e}$ decreasing. When TEM is dominant, $\upsilon_{\perp}=\upsilon_{E\times B}+\upsilon_{TEM}$ became more negative and the minimum Doppler shift could reach near $-1MHz$; when ITG is dominant, $\upsilon_{\perp}=\upsilon_{E\times B}+\upsilon_{ITG}$ became positive with maximum Doppler shift around $200kHz$. Transitions between TEM and ITG would make $\upsilon_{\perp}$ has different amplitudes in the positive and negative directions and then generate the ETRO harmonics on $\upsilon_{\perp}$ spectrum, as shown in Fig.\ref{f2}(b). To display the turbulence transition process clearly, panels (d) and (e) show the TEM and ITG intensity respectively through filtering different frequency components of the turbulence spectra, and the inverse growth and decline processes of TEM and ITG kept very well. Panel(f) gives particle flux $\Gamma_p$ at the strike point on divertor plate, showing four regular positive peaks above the averaged level, and these outward particle fluxes should be caused by the TEM bursts.

\begin{figure}
	\includegraphics[width=8.5cm,clip]{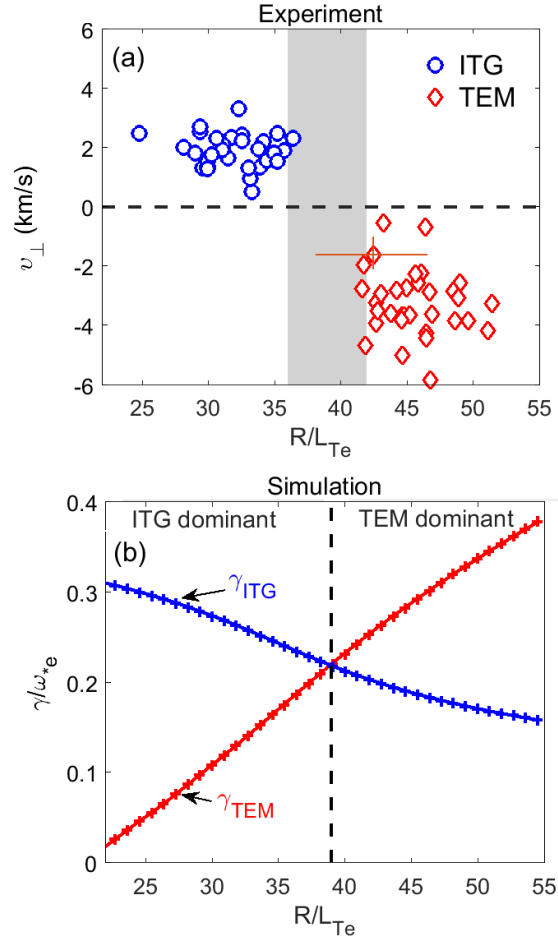}
	\caption{\label{f6} (a) $\upsilon_{\perp}$ versus $R/L_{T_e}$ during from experimental data; (b) the normalized linear growth rate ($\gamma/\omega^{*}_e$) of TEM and ITG versus $R/L_{T_e}$ from HD7 gyrokinetic simulation}
\end{figure}

Fig.\ref{f6}(a) plots the extreme values of $\upsilon_{\perp}$ as a function of $R/L_{T_e}$ during ETRO evolution. It could be found that the ITG-TEM transition threshold is about $36-42$. To further confirm the threshold, the toroidal gyrokinetic eigenvalue code HD7 \cite{Dong1995POP,Han2017NF} was used to calculate the linear growth rates of ITG and TEM with the following experimental parameters:  $R/L_{n_e}=R/L_{n_i}=9$, $R/L_{T_i}=20$. The results in Fig.\ref{f6}(b) showed that the linear growth rates of ITG and TEM have a crossover around $R/L_{T_e}=39$, consistent with the experimental results. Unlike the ITG-TEM transition reported previously \cite{Xiao2010PRL,Rice2011PRL,Zhong2013PRL}, here the transition could be periodly triggered back and forth to maintain a quasi-steady plasma confinement through the ETRO.

Based on the measurements from various diagnostics, these conclusions on the ETRO features during stationary I-mode on EAST could be drawn: 1) ETRO is an azimuthally symmetry and radially localized structure dominated by $T_e$ fluctuations; 2) the generation location of ETRO is close to the temperature pedestal top, with $10-30\%$ relative perturbation amplitude on $R/L_{T_e}$ due to the drastic outward decay; 3) ETRO is sustained by the alternative transition between TEM and ITG, and forms a self-organizing system together with turbulence and transport transitions. The process is probably the intrinsic mechanism to sustain the long-time stationary I-mode on EAST. 

It should be emphasized that ETRO should not be categorized as the limited cycle oscillation (LCO) phenomena, which have been widely reported during L-mode to H-mode transition \cite{Kim2003PRL}. ETRO frequency is $3-4$ times larger than LCO frequency on EAST \cite{XuPRL2011} and the most important point is that only different turbulence phase velocities could well explain the amplitude asymmetry of $\upsilon_{\perp}$ and then the ETRO harmonics. Another long-term conjecture on I-mode is that WCM during I-mode is probably the key player to drive outward flux and maintain particle transport \cite{Manz2015NF}, while in our experiments, WCM would synchronously grow with TEM and modulate TEM (the modulation could be directly seen in Figs.\ref{f5}(c) and (e)), as well as the corresponding transports, which is also consistent with these previous results.

This work was supported in part by the National MCF Energy R$\&$D Program under Grant Nos.2017YFE0301204 and 2018YFE0311200, Natural Science Foundation of China under Grant Nos.U1967206, 11975231 and 11922513.


\section*{References}
\bibliographystyle{prsty}
\bibliography{I-mode}

\end{document}